# A Framework for a High Throughput Screening Method to Assess Polymer/Plasticizer Miscibility

*Lois Smith[a], H. Ali Karimi-Varzaneh[b], Sebastian Finger[b], G. Giunta,[a,c] A. Troisi[d], Paola Carbone[*a]*

[a] Department of Chemical Engineering, School of Engineering, The University of Manchester, Oxford Road, M13 9PL, Manchester, United Kingdom

[b] Continental Reifen Deutschland GmbH, Jädekamp 30, D-30419 Hanover, Germany

[c] BASF, Carl-Bosch-Strasse 38, 67056, Ludwigshafen, Germany

[d] Department of Chemistry, Department of Chemistry, Crown Street, L69 7ZD, Liverpool, United Kingdom

[*]paola.carbone@manchester.ac.uk

**For Table of Contents use only:**

**Miscibility Predictors**

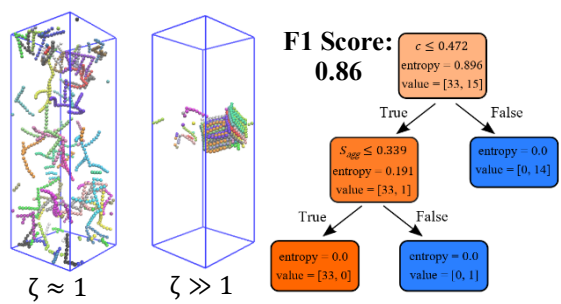

ABSTRACT Polymer composite materials require softening to reduce their glass transition temperature and improve processability. To this end, plasticizers, which are small organic molecules, are added to the polymer matrix. The miscibility of these plasticizers has a large impact on their effectiveness and therefore their interactions with the polymer matrix must be carefully considered. Many plasticizer characteristics, including their size, topology and flexibility, can impact their miscibility and, because of the exponentially large numbers of plasticizers, the current trial-and-error approach is very ineffective. In this work we show that using molecular simulations of a small dataset of 48 plasticizers, it is possible to identify topological and thermodynamic descriptors that are proxy for their miscibility. Using *ad-hoc* molecular dynamics simulation set-ups that are relatively computationally inexpensive, we establish correlations between the plasticizers' topology, internal flexibility, thermodynamics of aggregation and their degree of miscibility and use these descriptors to classify the molecules as miscible or immiscible. With all available data we also construct a decision tree model which achieves a F1 score of $0.86 \pm 0.01$ with repeated, stratified 5-fold cross-validation, indicating that this machine learning method is a promising route to fully automate the screening. By evaluating the individual performance of the descriptors, we show this procedure enables a 10-fold reduction of the test space and provides the basis for the development of workflows which can efficiently screen thousands of plasticizers with a variety of features.

**Introduction**

The development of rubber-based polymer composite materials has been significant within the manufacturing industry, with research tuning the processing and composition of these systems to improve their rheological and mechanical properties spanning decades. [1-5] It is now well established that the addition of solid particles, including carbon black [6-10] and silica [6, 7, 11, 12], can reinforce the rubber matrix resulting in polymer composites with greater strength, durability and decreased rolling resistance. [9, 10, 13] In addition to these fillers, small diluent molecules known as plasticizers (PLs) can be added to the rubber matrix to improve the filler/polymer adhesion properties, thus preventing filler aggregation within the polymer bulk. [14] Most commonly used within the tire industry are synthetic, petroleum-based PLs such as aromatic and paraffin oils that, when added to the polymer, normally polyisoprene, improve its mechanical properties and lower its glass transition temperature, $T_g$. [15-20]

A key feature for an effective PL is its miscibility within the polymer matrix. [21-23] This, however, is difficult to predict due to the complex interplay of enthalpy and entropy of mixing typical of polymer melts and the challenges encountered in experimentally characterizing the degree of mixing. This is usually done using indirect measurements such as the compound's viscosity or transition temperatures, [24, 25] or predicted through the calculation of the Hildebrand solubility parameter, δ, which is equal to the root-squared of the cohesive energy density (CED). In this case, the solubility is assessed via the calculation of $X_{12} = [\delta_1 - \delta_2]^2$ where 1 and 2 would, in this case, refer to the plasticizer and the polymer respectively. The higher the value of $X_{12}$, the more immiscible the plasticizers and polymers are. It is the norm to assume that if $X_{12} > 4$ MPa the two are immiscible. This method, while in principle useful, is plagued by problems including the fact that values of δ reported in literature have huge variability depending on the experimental methods used to measure CED and the effect of temperature on the experimental results.[26] Recently, the accuracy of the Hildebrand (and Hansen) methods has been quantitatively assessed and it has been shown that it is dependent on the solubility of the molecules and varies between 60% and 75% for apolar systems. [27]

Over the recent years, computer simulations have become a powerful and efficient tool to identify relevant PL properties, in part because they allow the study of model features systematically and in isolation. Several recent computational studies have attempted to link PL rigidity, chemistry, molecular weight and interaction strength with the polymer matrix to their miscibility.[28-33] Atomistic simulations are particularly useful when specific directional interactions such as hydrogen bonds play a role in the plasticization effect [29], however due to their high computational cost, they cannot be used for screening large families of molecules. To overcome this problem, coarse-grained models can be used as less computationally intensive alternatives. [30, 32, 34] These models have been successfully employed in polymer simulations to study structural features and plasticization effects [30, 35-37] as well as to study polymer blends.[38]

Recently, using a chemically-specific bead and spring model [33], we studied how the addition of small diluents affects a polymer/filler interface and which structural features improve their chances of adsorption onto a graphitic filler surface. There, we showed that just changing the backbone rigidity of the additives affects not only their degree of dispersion in the polymer matrix but also their adsorption on the surface of the filler. PLs with a flexible backbone (akin to short oligomers) remained evenly dispersed throughout the system while those with rigid backbone (akin to olefins with high degree of unsaturation) either formed clusters in the matrix or became adsorbed to the filler surface. The results indicated that clustering of long, rigid PLs, and the solubility of flexible PLs, were phenomena occurring in polyolefins irrespective of the polymer chemistry and molecular weight. This work highlighted the importance of the PL's molecular weight and rigidity, two features that should be captured by any virtual screening approach.

Even when using coarse-grained models, individual studies exploring the relationship between PLs' chemical and structural properties and their miscibility within a polymer matrix can be computationally time consuming. This cost is greatly compounded the larger the parameter space is made. At the moment, the additives manufactured for commercial use display a wide range of properties including varying chemical compositions and molecular weights. In order to minimize the number of experiments when choosing the optimal additives, it is therefore important that a robust structure-property relationship is established *a-priory*, particularly since recent sustainability concerns are motivating the replacement of conventional PLs with bioderived ones. [31, 39-42]

To avoid carrying out computationally expensive calculations, one could identify molecular descriptors that correlate with the molecule miscibility thus working as proxies. With the increase in popularity of machine learning methods, such correlations can now be established using, if available, large and well-curated databases. Molecular modelling can help in building or enriching such databases when the experimental data are sparse or unreliable. [43, 44] However, rather than performing hundreds of simulations, one could use machine learning approaches to reduce the parameter space and identify potential material candidates for the application of interest. [45-47] When the parameter space is large, this screening and pruning of data can, however, be complicated. For example, Jablonka *et al.* [48] in the search for the optimal block copolymers to stabilize colloidal dispersions, used a modified version of the ϵ-PAL algorithm [49, 50] to reduce polymer search space by calculating the set of optimal descriptors that form a Pareto front. Using a coarse-grained model of copolymers representative of dispersants used in solid suspensions, their descriptors (or performance indicators) were the adsorption free energy onto a surface, the dimer free energy barrier between two polymers and the polymer radius of gyration $R_g$. These properties characterise polymer/surface adhesive strength, polymer/polymer repulsion and polymer viscosity, all of which are criteria to consider when selecting an optimal polymer to prevent the flocculation of suspended particles.

The scope of the present work is to use molecular simulations to identify molecular or thermodynamic descriptors for the prediction of the solubility of medium sized organic molecules in polymer bulks so that they can act as PLs. Such descriptors should be quickly computable from simplified simulation set-ups, such as in vacuum, and ideally experimentally measurable. As in the work by Jablonka *et al.*, we employ a coarse grained model for both the polymers and the medium sized molecules following our previous work.[33, 34] For polyolefins the major contributor to the solubility is related to the geometry of the chain and local density.[26] Thus, the chosen coarse-grained model, whose parameters have been developed fitting the experimental packing length among other geometrical chain properties,[51] should provide results (in terms of solubility) comparable with an all-atom model. It is important to highlight however, that such coarse-grained model might not be sufficiently accurate if more specific interactions (such as hydrogen bonds or hydrophobic/hydrophilic interactions) drive the miscibility.

Performing long Molecular Dynamics (MD) simulations of 48 polymer/PL systems, we identify three descriptors, two geometrical and the other thermodynamic, that can be quickly calculated, and demonstrate that these can be used as simple proxies to rapidly screen for potentially miscible molecules. The procedure is fully automated, including the method to quantitatively establish whether phase separation had occurred, thus opening up the possibility to efficiently perform high throughput simulations to identify candidate molecules on which to carry out atomistic simulations or experiments.

## Methodology

*Coarse-grained model*

To construct the polymer/PL systems we followed our previous work.[34] We used the coarse-grained model developed by Svaneborg and co-authors [51] of a cis-(1,4)-polyisoprene (PI) polymer matrix filled with low molecular weight PI chains which act as PLs. Despite the level of coarse-graining, this model has been proven successful in reproducing structural and mechanical behaviour of PI and several other polyolefins [51] and predictions of the adsorption of plasticizers with different molecular rigidity onto filler surfaces has been validated experimentally.[22] The PI beads interact via the Weeks-Chandler-Andersen (WCA) potential which is purely repulsive and equivalent to the standard 12-6 Lennard-Jones (LJ) potential shifted to zero and with its attractive tail cut off, given by the following expression

$$U_{WCA}(r) = \begin{cases} 4\epsilon\left[\left(\frac{\sigma}{r}\right)^{12} - \left(\frac{\sigma}{r}\right)^{6} + \frac{1}{4}\right] & \text{for } r < 2^{1/6}\sigma \\ 0\epsilon\left[\left(\frac{\sigma}{r}\right)^{12} - \left(\frac{\sigma}{r}\right)^{6} + \frac{1}{4}\right] & \text{for } r > 2^{1/6}\sigma \end{cases} \quad (1)$$

where $\epsilon = k_B T = 2.477$ kJ mol$^{-1}$ is the potential well depth and $\sigma = 0.4136$ nm is the effective diameter of the bead, with each bead approximately 67% of the total mass of a PI monomer, $m_b = 45.62$ g mol$^{-1}$. This value of bead mass is such that the model is able to reproduce the correct Kuhn length and Kuhn segment density of PI.

For the bonded interactions we used the finite-extensible non-linear elastic (FENE) potential

$$U_{\text{FENE}}(r) = -\frac{kR_0^2}{2}\ln\left[1 - \left(\frac{r}{R_0}\right)\right] \quad (2)$$

where $k = 30\epsilon\sigma^2 = 434.4$ kJ mol$^{-1}$ nm$^{-2}$ is a force constant related to the bond strength of the inter-bead interactions and $R_0 = 1.5\sigma = 0.6204$ nm is the maximum bond length. The sum of the potentials given by Equation ( 1 ) and Equation ( 2 ) is anharmonic with a minimum that sets an equilibrium bond length of $0.965\sigma = 0.3991$ nm at a temperature of $T = 298.1$ K bead density of $\rho_b = 0.85\sigma^{-3} = 12.01$ nm$^{-3}$.

Owing to the coarse-grained nature of the model, its computational efficiency is high. This allows long simulation times at a reduced computational cost, allowing us to access a range of dynamical and thermodynamic properties with long relaxation times. To control the degree of flexibility to the PI chains an extra bending potential was introduced, [52]

$$U_{bend}(\theta) = \kappa(1 - \cos\theta) \qquad (3)$$

where $\theta$ is the bond angle between sets of three successive bonded beads and $\kappa = 0.206\epsilon = 0.5103$ kJ mol$^{-1}$ is a stiffness parameter which is chosen such that the correct number of Kuhn segments per Kuhn volume of PI is obtained. [51] The flexible PLs simulated in this work use the same interactions throughout the molecule i.e. the backbone and side chains have equal flexibility and interact with the PI matrix in the same way. We further simulated rigid, rod-like PLs, for which the aforementioned bending potential was replaced by a harmonic potential

$$U_{rod}(\theta) = \frac{1}{2}k_h(\theta - \theta_0)^2 \qquad (4)$$

where $k_h = 100$ kJ mol$^{-1}$ rad$^{-2}$ is a force constant and $\theta_0$ is an equilibrium bond angle which is fixed according to the geometry of each PL molecule (see Figure 1). This model has been used in previous studies of polymer composites. [33, 53, 54] All simulations in this work were performed using the GROMACS 5.0 Molecular Dynamics (MD) simulation package. [55]

*Plasticizers' Topology*

In this work, we chose a set of 48 PLs which sample the wide range of topological properties typical of these molecules. For all topologies, the backbone length was fixed at 10 beads and the side chain length, $L_{side}$, and the side chain grafting density, $\rho_{side}$, were varied, the latter defined as the ratio of number of side chains to backbone beads. The backbone length of 10 beads was chosen based on previous results [33] which showed 10 beads is sufficient length for rod-like PLs, without the presence of side chains, to start forming agglomerates. This length also allows for the selection of a range of desirable values of $\rho_{side}$. The values of $L_{side}$ varied from 3 to 9 beads and $\rho_{side}$ ranged from 0.2 to 0.5 in intervals of 0.1. The side chains are all evenly spaced beginning from the first bead of the backbone. For each unique PL topology with $L_{side}$ values of 3-7 beads, we modelled one molecule as flexible, using the bending potential in Equation 3, and one as rod-like with the potential given in Equation 4. For the latter, we fixed the equilibrium bond angles within the PLs, as displayed in Figure 1. The resulting number of PLs simulated was 40 (2 × 20). In each case, the bending potential was applied to the whole PL molecule *i.e.* both the backbone and side chain angles. It is important to notice that the nomenclature used to describe the PL topology (*i.e.* number of backbone and side chain beads) applies only to the rod-like molecules, where the side chains are restrained in place by the equilibrium bond angles (see Figure 1). Due to the nature of the bending potential imposed on flexible PLs, a side chain placed on the first or last bead of the backbone is an effective extension to the 10 beads in our definition. We find that simulating flexible PLs with their side chains present on neither the first nor the last beads of the PL backbone makes no practical difference to our results (see *Further Flexible PL Simulations* in the Supporting Information (SI)). For the remaining 8 PL topologies ($L_{side}$ = 8, 9), we modelled only rod-like molecules as, from previous work [33, 34], we anticipate that flexible or low molecular weight PLs will remain miscible in the PI matrix and so the higher molecular weight, rod-like molecules produce a more balanced dataset.

To more easily distinguish between PL topologies, we have produced an alphanumeric code which will be referred to from now on within this work, an example is given in Figure 1.

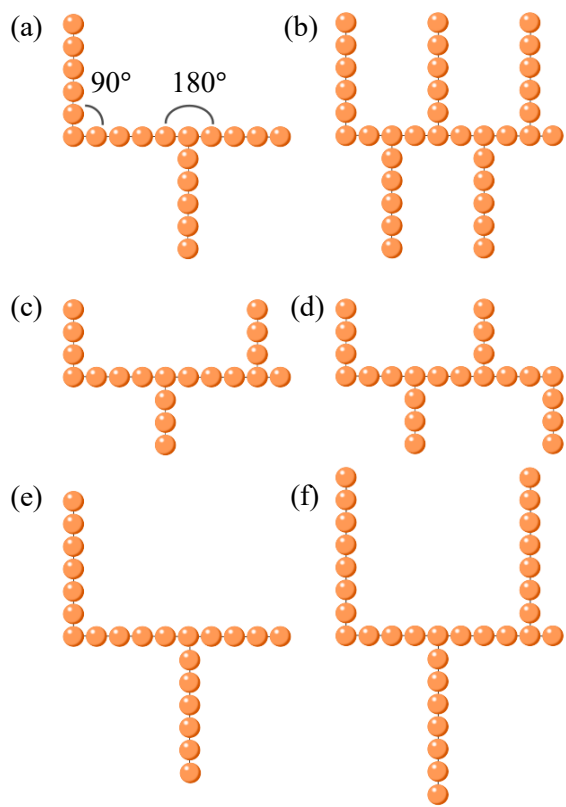

**Figure 1.** Examples of the PL topologies simulated in this work. The code is as follows: (backbone length)B-($L_{side}$)S-($\rho_{side}$)-r/f, where the character "r" means "rod-like" and "f" means "flexible". The topologies shown are **(a)** 10B-5S-0.2-r/f, **(b)** 10B-5S-0.5-r/f, **(c)** 10B-3S-0.3-r/f, **(d)** 10B-3S-0.4-r/f, **(e)** 10B-6S-0.2-r/f and **(f)** 10B-7S-0.3-r/f. Equilibrium bond angles applied to the rod-like PLs are labelled in **(a)**.

*Quantifying Miscibility*

To study PL miscibility, we performed the following simulations. Firstly, we randomly inserted PL molecules and 72 PI chains of length 300 beads into a simulation box of size 8.575 × 8.575 × 25 nm. The length of the PI chains was chosen to be above the entanglement molecular weight and the number of PLs such that the concentration of PL beads was approximately 5 *phr* (per hundred rubber), commonly used in industry [56], where $phr = \frac{N_{PL}}{N_{PI}} \times 100$ and $N_{PL}$ and $N_{PI}$ are the numbers of PL and PI beads respectively. We elected to only study one PI chain length based on our previous work using the same model [33], in which PI lengths of 1, 10, 150 and 300 were investigated. It was found that, for each PL molecule studied, the miscibility was consistent regardless of PI length, with the exception of the PI solvent (i.e. a monomer solution) for which all PLs studied were miscible. For this reason, we kept the PI length consistent throughout our simulations. The full list of simulated systems is reported in Table S1 of the SI.

After energy minimisation, to equilibrate the density of the box, we performed a 50 ns simulation in the isobaric-isothermal (NPT) ensemble at 298.1 K and 2830.87 bar. Such high pressure is needed to reproduce approximately the experimental density of PI of 910 kg m$^{-3}$. [57] The pressure was controlled by the Berendsen barostat with a time constant of $\tau_p = 2$ ps and the system evolved through Langevin dynamics, which handles the system temperature, with a friction coefficient of $\Gamma = m_b \tau^{-1} = 12.85$ g mol$^{-1}$ ps$^{-1}$, where $\tau = \sigma(m_b \epsilon^{-1})^{0.5}$. Equations of motion were integrated with a time step of $0.01\tau = 0.017$ ps [58] and periodic boundary conditions were applied in all directions. The final configuration from the NPT simulation was then used as the starting configuration for a production run in the canonical (NVT) ensemble. In order to quantify the PL miscibility, we followed the procedure from our previous work. [33] As a brief summary, we calculated the integral of the PL-PL centre of mass radial distribution function ($g(r)$) between $r = 0$ nm and $r = r_{cut} = 3$ nm, $N = 4\pi \int_0^{r_{cut}} r^2 g(r) \rho_{tot} \, dr$, where $\rho_{tot}$ is the PL number density of the whole system. [33] The average PL number density within this region is then $\rho_{in} = (N + 1)/(\frac{4}{3}\pi r_{cut}^3)$. Thus, we define the miscibility parameter, $\zeta$, as the ratio of $\rho_{in}$ with the average PL number density in the remaining bulk material, $\rho_{out} = (N_{tot} - N_{in} + 1)/(V_{tot} - N_{in})$.

ζ is then

$$\zeta = \frac{4\pi \int_0^{3\text{nm}} r^2 g(r) \rho_{tot}\, dr + 1}{\rho_{out} V_s} = \frac{\rho_{in}}{\rho_{out}} \qquad (5)$$

where $V_s = \frac{4}{3}\pi r_{cut}^3$. ζ is a versatile descriptor for miscibility since it can be used for systems of different sizes and chemistries, with consistently accurate results. [33] Values of ζ were recorded every 2000 ps and through a trajectory of sufficient length such that equilibrium PL clustering was observed. The average value of ζ was then extracted over the last 1 $\mu$s of the trajectory. According to our definition, ζ equals 1 for a completely evenly dispersed system and increases in value with decreasing PL miscibility.

*Geometric Descriptors*

We identified two geometrical descriptors that correlate with PL miscibility: the PL square radius of gyration, $R_g^2$, and the acylindricity, $c$, which are commonly used to analyse instantaneous structural features of polymers. [59, 60] Both can be calculated from the eigenvalues, $\lambda^2$, of the gyration tensor, $S$, and, as we are going to show, directly from simulations performed in vacuum, thus greatly decreasing the computational cost of the simulation when compared to simulating a PL molecule within the PI melt.

The components of the gyration tensor, $S_{ij}$, are given by the following equation

$$S_{ij} = \frac{\sqrt{\sum_k m_k (x_{ik} - x_{jk})^2}}{M} \qquad (6)$$

where $m_k$ is particle mass, $x_{ik}$ and $x_{jk}$ are the *i-th* and *j-th* components of the position vector of the *k-th* ~~molecule~~ bead respectively, and $M$ is the total mass of the ~~system~~ PL molecule. The eigenvalues of $S$, $\lambda_1^2 \geq \lambda_2^2 \geq \lambda_3^2$, are the principal moments of $S$, and represent the characteristic lengths of the ellipsoid describing the molecule. The square radius of gyration, $R_g^2$ is then calculated as

$$R_g^2 = \lambda_1^2 + \lambda_2^2 + \lambda_3^2 \qquad (7)$$

while the acylindricity, $c$, which describes the deviation the PL molecule from cylindrical symmetry, is calculated as [61]

$$c = \lambda_2^2 - \lambda_3^2 \qquad (8)$$

where higher values of $c$ represent a greater departure from cylindrical symmetry. Also tested was the PL relative shape anisotropy, $\kappa^2$, which is a dimensionless quantity, ranging from 0 to 1, describing the spherical symmetry of the molecule [61]

$$\kappa^2 = \frac{3(\lambda_1^4 + \lambda_2^4 + \lambda_3^4)}{2(\lambda_1^2 + \lambda_2^2 + \lambda_3^2)^2} - \frac{1}{2} \qquad (9)$$

We found, however, that this descriptor is ineffective at distinguishing between miscible and immiscible PL molecules so it was excluded from the screening procedure.

These descriptors represent the instantaneous conformational properties of the PL molecules, allowing us to numerically quantify information about their size and shape. To verify whether these properties are sufficiently accurate if determined from a very rapid simulation of the PL in vacuum, thus saving significant simulation time, we performed simulations of a single PL molecule both in vacuum and dissolved in a PI melt. We equilibrated the density of these systems with a 50 ns NPT simulation, followed by a further 800 ns NVT simulation to ensure the PI mean square internal distance plateaued. Results were then extracted over a 2 μs production run for good statistics. For these tests, we chose a small subset of 8 PL topologies. We found that, in the case of rod-like PL molecules, the values calculated in vacuum and in the PI melt were almost identical. For flexible PL molecules, there is a more noticeable difference between the value as the lack of a PI melt in the vacuum simulations allows the PL molecules to swell, increasing their effective size. However, the overall trend across the PL topologies is similar across the two simulation types tested and considering the advantageous computational efficiency of vacuum simulations over simulations with a PI melt, we choose to calculate these descriptors for the remaining PLs only in vacuum. The results for these tests are displayed in Figure S1 of the SI.

*Configurational Entropy*

The WCA potential used to model the non-bonded interactions is purely repulsive which results in entropically driven PL miscibility behaviour. This means we can exclude enthalpy as a potential driver for PL clustering. Considering this, we evaluated an approximate aggregation entropy, $S_{agg}$, as

$$S_{agg} = S_{PL/PL} - S_{PL} \qquad (10)$$

where $S_{PL/PL}$ is the entropy of a PL molecule within a cluster of other PLs and $S_{PL}$ is the entropy of an isolated PL. The difference between the two values should be indictive of whether it is entropically favourable for the PL molecules to form clusters or to dissolve in the PI matrix. Thus, a positive value of $S_{agg}$ indicates that PL/PL aggregation is favoured, and vice versa for a negative value. In this work, we only consider the PL configurational entropy to calculate the aggregation entropy given by the following Gibbs definition

$$S_{config} = -k_B \sum_{i}^{N_c} p_i \ln(p_i) \qquad (11)$$

where $S_{config}$ is the configurational entropy, $N_c$ is the total number of configurations a PL acquires and $p_i$ is the corresponding probability that each configuration will occur. In our case, this translates to calculating $S_{config}$ for a PL in a cluster of other PLs ($S_{PL/PL}$) and that of a PL in isolation ($S_{PL}$). To calculate $S_{config}$ from the simulations, $p_i$ is the probability associated with the probability distribution of the PL end-to-end distance ($R_{ee}$) following a previous work. [62] The choice of this internal coordinate only partially accounts for the shape of the PL molecule. For our purposes, however, this is sufficient to capture the relative behaviour of the PL molecule miscibility.

As the aim is to screen a large number of hypothetical PLs, we verified whether $S_{PL}$ can be calculated from vacuum simulations rather than from simulations of a single PL immersed in the PI melt. To test this, we again performed a comparative analysis between the two types of simulations on a subset of 8 PL topologies; the results for which are reported in Figure S2 of the SI. We found that the presence of PI chains does not significantly impact the results, thus the calculation of $S_{PL}$ for the remaining PLs can be done using vacuum simulations.

A second system of a box of only PLs molecules to mimic a PL cluster was used to calculate $S_{PL/PL}$. We performed a 50 ns NPT equilibration with the same interaction parameters and pressure as described in the previous sections in order to ensure that all systems are in consistent thermodynamic conditions. The number of PL beads in the box was fixed to approximately 6500, this ensures each simulation box is approximately 8 nm × 8 nm × 8 nm post NPT equilibration which is large enough to avoid finite size effects yet still avoids very computationally expensive simulations. Finally, we performed an NVT production run for 5 − 15 μs, until the average entropy reached a plateau over 600 ns blocks of the simulation trajectory. The time to reach equilibrium varied vastly between systems, with rod-like PLs showing the longest relaxation time. Further details of the equilibration of these systems is given in Section *PL/PL System Equilibration* of the SI. It is worth noticing that the entropy descriptor itself only accounts for the change in conformational entropy of a plasticizer being in a cluster compared to be in solution. Thus, not included in our considerations are entropies such as translational or rotational and additionally the conformational entropy changes of the polyisoprene chains surrounding the plasticizers. The latter is however small, and it scales with the inverse of the number of monomers.[26]

*Decision Tree Method*

In this work, we choose to evaluate the performance of each PL descriptor individually, with an *ad hoc* procedure outlined in the *Results* section. This was done in order to easily assess the performance of each descriptor on our relatively small amount of data. Considering the future scalability of this work, however, it is useful to demonstrate with this small dataset already that more conventional methods for automatically selecting descriptor thresholds are also effective. Performing feature selection to identify the combination of features that best map to PL miscibility can also aid in reducing the dimensionality of future data collection. With this in mind, we performed a decision tree analysis. This technique, which determines a design path for selecting miscible PLs through a series of branching nodes, can effectively be applied to binary classification tasks in high throughput screening of polymers,[63] as in this work, and also in regression tasks for polymer property prediction with relatively small data sets. [64]

We constructed the tree using the ML library Scikit-Learn,[65] which uses an optimised version of the CART algorithm, implemented in Python 3.11.4. Node splitting was determined by evaluating the data entropy at each node. The performance of the model was evaluated using repeated 5-fold cross-validation over 100 iterations and folds were selected using stratified random sampling to avoid bias in the dataset, which decision trees are prone to.

## Results

*Miscibility Factor, $\zeta$*

In this section, we report the findings of the full PL/PI simulations. The miscibility parameter, $\zeta$, calculated for all 48 PL topologies are given in Figure 2. The results show that all systems with flexible PLs have a high level of miscibility, as every value of $\zeta$ is close to 1. This is consistent with previous simulations [22, 33, 66] and the experimental findings of Lindemann and co-workers [22] showing that PLs with higher backbone flexibility show lower tendency to cluster in the polymer matrix when compared to less flexible PLs.

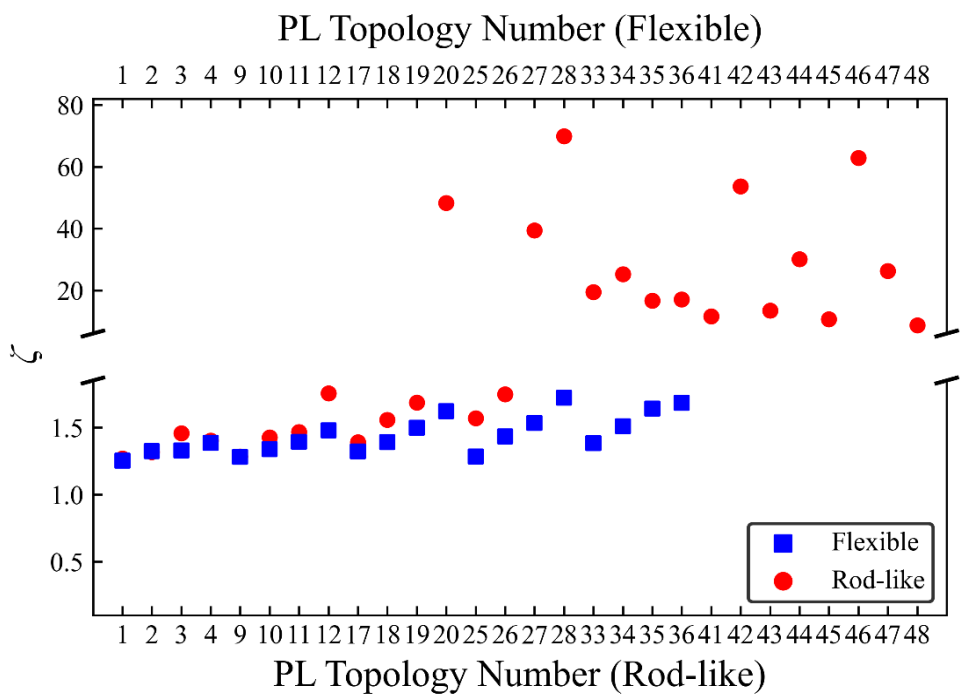

**Figure 2**: Miscibility parameter, ζ, against PL topology number for the PL/PI simulations for flexible (blue square) and rod-like (red circle) PL topologies. The PL topology number on the bottom axis has the same initial geometry as the corresponding topology number on the top axis. The error bars are the standard errors of the data are smaller than the point size. Note the difference in scale between y-axis breaks. A full list of PL topology numbers and their corresponding alphanumeric code can be found in Table 1 of the SI.

The rod-like PLs on the other hand, show a wide variability in their miscibility depending on the value of $L_{side}$ and grafting density. Their average ζ values vary between 1.25 and 79.41 which is a result of different degrees of packing within the PL clusters of the different systems. We observe that, similarly to flexible PLs, some rod-like PLs remain miscible within the PI matrix, for example PL 10B-3S-0.4-r (labelled in Figure 2) with 4 side chains, each 3 beads in length. Despite this, the backbone of this PL exceeds the critical number for clustering in rod-like PLs. [33] This indicates that the presence of side chains impact the ability of the PLs to cluster. Example snapshots and ζ distributions of systems displaying different levels of PL clustering are displayed in Figures S4 and S5 of the SI. Among the rod-like PLs only, there are also some that show to fall within a meta-stable region in which PL clusters form and dissolve throughout the simulation. An example is topology 10B-6S-0.2-r, whose ζ values across the simulation oscillate between approximately 1.25 and 2.25, as shown in Figure 3. On the contrary, PL topology 10B-5S-0.2-f shows no meta-stable behaviour, as the ζ values remain close to 1 throughout the whole simulation with no appreciable cluster lifetime visible.

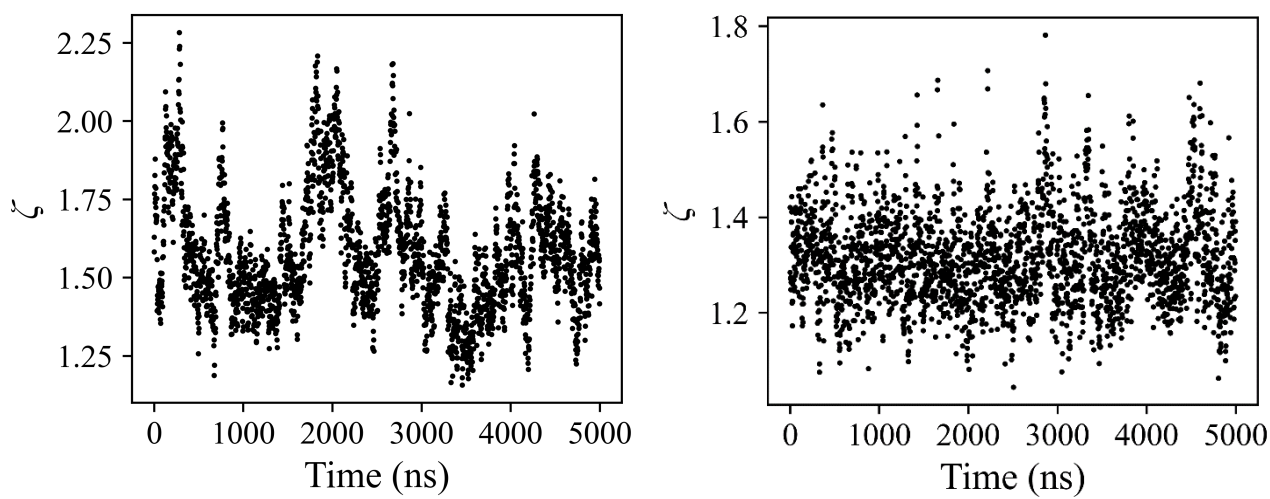

**Figure 3:** Miscibility parameter, ζ, against simulation time for PL topologies 10B-6S-0.2-r (left), displaying meta-stable behaviour, and 10B-5S-0.2-f, displaying a lack of meta-stable behaviour, immersed in a PI melt of 72 chains with molecular weight 300, at a PL concentration of approximately 5 *phr*.

The origin of this behaviour can be explained by considering the free energy of the system. If small and local variations in composition lead to an increase in free energy, the mixture can be considered to be in a meta-stable state in which PLs oscillate between aggregation and solvation. This is consistent with the theory of spinodal decomposition in polymer blends. [67]

Thus the oscillation of the ζ values along with maximum values achieved, can be used as an indicator for the formation of small and relatively short lived clusters and then to classify systems showing this behaviour as "meta-stable". We also notice that, the highest values of ζ achieved for these meta-stable systems are low relative to the those calculated for systems which display fully immiscible behaviour. In this work, we conservatively choose a cut-off value of $ζ = 2.7$ to distinguish between miscible and immiscible PL behaviour, a snapshot of which is shown in Figure S4 of the SI. Since the PL topologies we classify as meta-stable do not fluctuate near this range, we can take their average ζ value as a reliable indicator of whether the PL is miscible or immiscible within the PI melt.

*Correlation Between Geometric Descriptors and Miscibility*

Figures 4(a) and (b) show how the radius of gyration and the acylindicity of the PLs correlate with the value of ζ. The results show that for most of the PLs there is a correlation between the geometrical descriptors and the PL miscibility. In general, PLs with smaller $R_g^2$ and a lower deviation from cylindrical symmetry ($c$) are more likely to remain miscible within the PI matrix. This result indicates that a fast (the simulations have been run in vacuum) initial screening can by done just using these geometrical parameters, massively reducing the run time in simulations to determine PL miscibility. However, Figure 4 shows that there is a region (shaded) in the values of both $R_g^2$ and $c$ for which the correlation is weakened. For the PLs characterized by these of $R_g^2$ and $c$ values, it is not possible to determine whether they will mix or demix just using these geometrical descriptors. These PL topologies need therefore to be screened via the configurational entropy descriptor, whose results are presented in the following section.

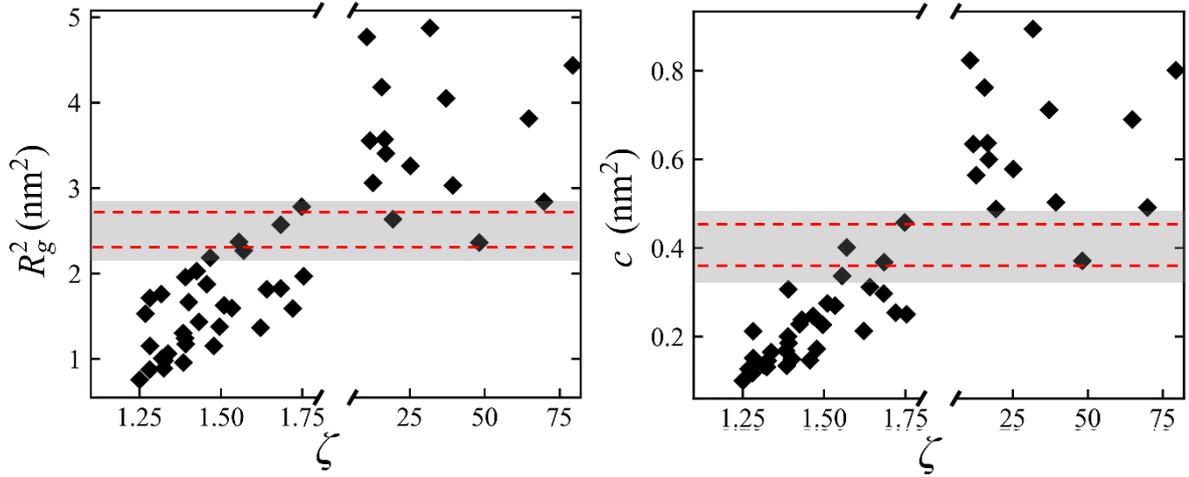

**Figure 4**: PL $R_g^2$ (left) and $c$ (right) against miscibility factor, ζ. The shaded region indicates a region of overlap between types of behaviour. The region, defined by the dotted lines, was chosen based on the error bars of $R_g^2$ and $c$, which are calculated with the standard error of block averages, and are smaller than the symbols on the graph. To be conservative, each bound of the shaded region was expanded by 40%.

*Correlation Between Aggregation Entropy and Miscibility*

Plotting $S_{agg}$ against ζ for the all the PL topologies, shown in Figure 5, we can again observe a separation in the values of $S_{agg}$ depending on the miscibility of the PLs. Value of $S_{agg}$ around zero correspond to PLs that have no entropic preference, in terms of their configurational entropy, to form clusters thus should remain dissolved. The higher the value $S_{agg}$ is, the larger the PL configurational entropy in the PL cluster is compared to that of the isolated molecule. Thus PLs with high positive value of $S_{agg}$ should cluster and demix. This correlation is well reflected in the results of Figure 5 which confirm the viability of $S_{agg}$ as a second level of screening. However, as for the geometrical descriptors, there is a region of overlap (shaded) where it is impossible to distinguish, via the value of configurational entropy difference, between miscible and immiscible PLs. For these cases full PL/PI simulations need to be carried out.

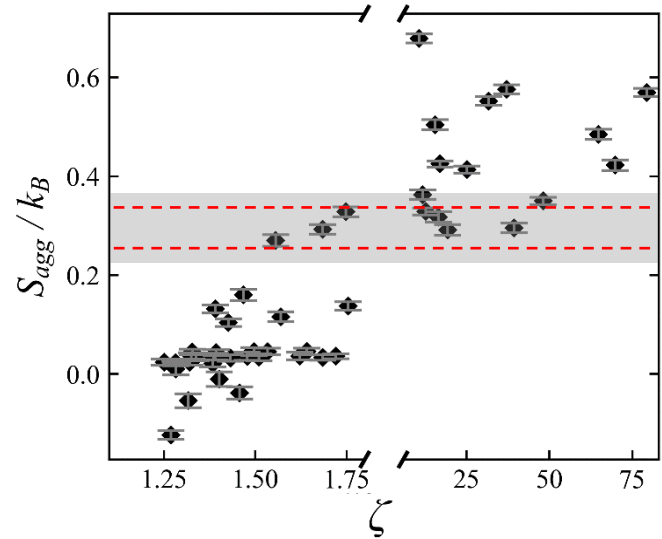

**Figure 5:** Difference in entropy between a PL in a PL/PL system and that of one in a vacuum, $S_{agg}$, against miscibility factor, $\zeta$, for each PL topology simulated. The shaded region, determined by the error bar overlap, signifies the area of overlap for which PL miscibility behaviour cannot be predicted with this descriptor and is marked with the dotted lines. To be conservative, each bound of the shaded region was expanded by 40%.

*Decision Tree Analysis*

Until this point, the boundaries of the overlapping regions (shaded areas in Figures 4 and 5) have been determined manually by observing the data. The method of selecting miscible/immiscible PLs is then to evaluate their descriptor values against each plot (Figures 4 and 5). In contrast, a decision tree can accept all the data and automatically perform efficient feature selection to determine a design path for miscible/immiscible PLs. It also eliminates any subjectivity induced by manually selecting the shaded regions, improving the reliability of the selection process. In our case, despite the limited amount of data available, as our descriptors are physics-based and scale well with miscibility, we are able to build a simple and efficient model. The decision tree is displayed in Figure 6. The majority of the classification is performed by the acylindricity descriptor, $c$, with the final node split classifying only one PL with the aggregation entropy descriptor, $S_{agg}$, which may be due to the level of class imbalance within our small dataset. To measure model performance, we carried out repeated stratified 5-fold cross-validation over 100 iterations and calculated its average F1 score, $0.86 \pm 0.01$. This is the harmonic mean of the model's precision and recall and provides an indication of well it can minimise false positive (precision) and false negative (recall) predictions. A value of 1 has both perfect precision and recall. From this, we can assume our model will generalise well to unseen data, although we point out that due to the small size of our dataset, the error on model predictive accuracy is possibly high [68] and only an expanded dataset will be able to mitigate this issue. Despite this, considering their good correlation with PL miscibility, the descriptors already identified are promising candidates for a decision tree model. Additionally, as previous simulations [33] have shown, the PL miscibility behaviour is driven by PL molecular architecture and not by the polymer matrix molecular weight or chemical composition; therefore we expect that similar threshold values to determine PL classification will be valid if a different polymer matrix is used.

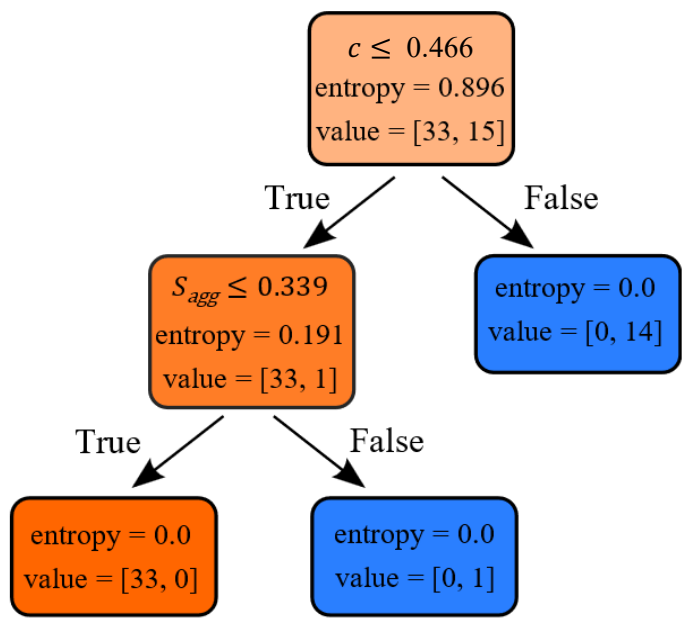

**Figure 6**: Decision Tree model to select miscible/immiscible PLs. Colour represents the majority classification, miscible (orange) or immiscible (blue), and opacity represents node purity. Each node is labelled with the condition for the subsequent node split, data entropy and the number of PLs of each classification ([miscible, immiscible]). We set the maximum depth of each model to 2, beyond which we saw no significant improvement in average F1 score.

*Discussion*

As observed in the case of the geometric descriptors, some PL features seem to be always associated to mixing or demixing. The flexible PLs are always miscible, irrespective of their topology. In contrast, the rod-like PLs have values of entropy that span a much larger range, suggesting a more complex behaviour. In general, PLs with shorter side chains and smaller values of $\rho_{side}$ display lower values of aggregation entropy, while PLs for which neither descriptors can predict the degree of miscibility (those in the shaded region in Figures 3 and 4) are all rod-like and are characterised by comparatively larger values of $L_{side}$ and $\rho_{side}$.

To identify emerging design rules, the values of $L_{side}$, $\rho_{side}$ and $\zeta$ are plotted for all 28 rod-like PL topologies in Figure 7. From our results, molecule flexibility plays the most important role in determining the PL miscibility, with all flexible PLs in the miscible range (see Figure 2), for this reason we only show the results for the rod-like PLs. In this case, there is a cooperative effect in the PL $L_{side}$ and grafting densities. For example, configuration 10B-3S-0.3-r ($L_{side} = 3$) is miscible within the PI matrix but increasing $L_{side}$, for example in topology 10B-7S-0.3-r ($L_{side} = 7$), results in immiscible behaviour. Conversely, PLs with a shorter value of $L_{side}$, for example in topology 10B-5S-0.5-r, can become immiscible if $\rho_{side}$ is increased. It is interesting to notice that the correlation between PL topology and miscibility that emerges from our analysis qualitatively agrees with the predictions obtained with the available Hildebrand solubility parameters. Observing the trend in $\delta$ values across a small series of short hydrocarbons (akin to our plasticizer molecules) with varied number of side chain groups, assuming $\delta$ for polyisoprene to be 16.77 MPa$^{1/2}$,[69] the available data seems to indicate that, in agreement with our predictions, saturated hydrocarbons between C5 (pentane) and C8 (octane) would dissolve ($X_{12} < 4$ MPa). Moreover, the data shows that the addition of side methyl side groups increases from $X_{12}=2.16$ for n-heptane to $X_{12} = 3.5$ and 5.8 for 2,3-dimethyl pentane and 2,4-dimethyl pentane respectively, indicating, as our simulation results, that the addition of side group can hinder solubility.[70]

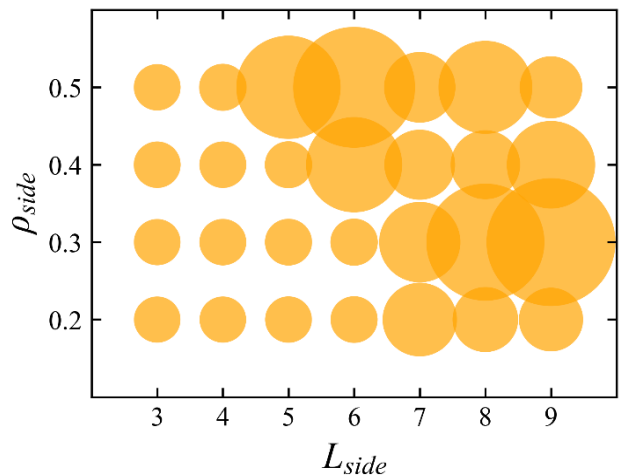

**Figure 7**: Side chain grafting density, $\rho_{side}$, versus side chain length, $L_{side}$, for the rod-like PL topologies. The size of the circles scales linearly with the value of the miscibility factor, ζ, therefore larger points represent less miscible PLs.

We summarize the performance of the task of classifying the PL into miscible/immiscible on the basis of single easily computable parameters in Figure 8. Here, we assign each PL topology a circle cut into thirds, with each portion representing one of the 3 screening procedures: $R_g^2$, $c$ or $S_{agg}$. If the descriptor correctly predicts miscible/immiscible behaviour, the portions are colored green/red respectively and if the behaviour cannot be predicted by the descriptor, the portion is left blank. Figure 8(a) shows that plot for each combination of $L_{side}$ and $\rho_{side}$ for the rod-like PLs for which the descriptors have varying success in identifying the correct miscibility behaviour. The behaviour of all the flexible PLs is always correctly identified by all 3 descriptors and as such are not pictured.

From our description, a corresponding score is then given to each PL topology, details for which are displayed in Figure 8(b). If a descriptor can correctly identify a PL as miscible, a value of +1 is added to its score and if, conversely, the descriptor correctly identifies immiscible behaviour, a value of −1 is added to its score. PLs can achieve a 'perfect' score if all three descriptors correctly predict miscible (score = 3/3) or immiscible (score = −3/3) behaviour. If a descriptor is unable to determine PL behaviour (those in the shaded regions of Figure 4 and Figure 5), a value of +0 is added to its score. A histogram of these results for both the rod-like and flexible PLs is displayed in Figure 8(c), showing the number of PLs that achieve each score. From this, we can see that, among the miscible PL topologies (represented by positive scores), the majority of them are correctly predicted by all 3 descriptors. Among immiscble PL topologies (represented by negative scores), the descriptors are generally less accurate but the correct behaviour can be identified by at least 1 descriptor in the majority of cases.

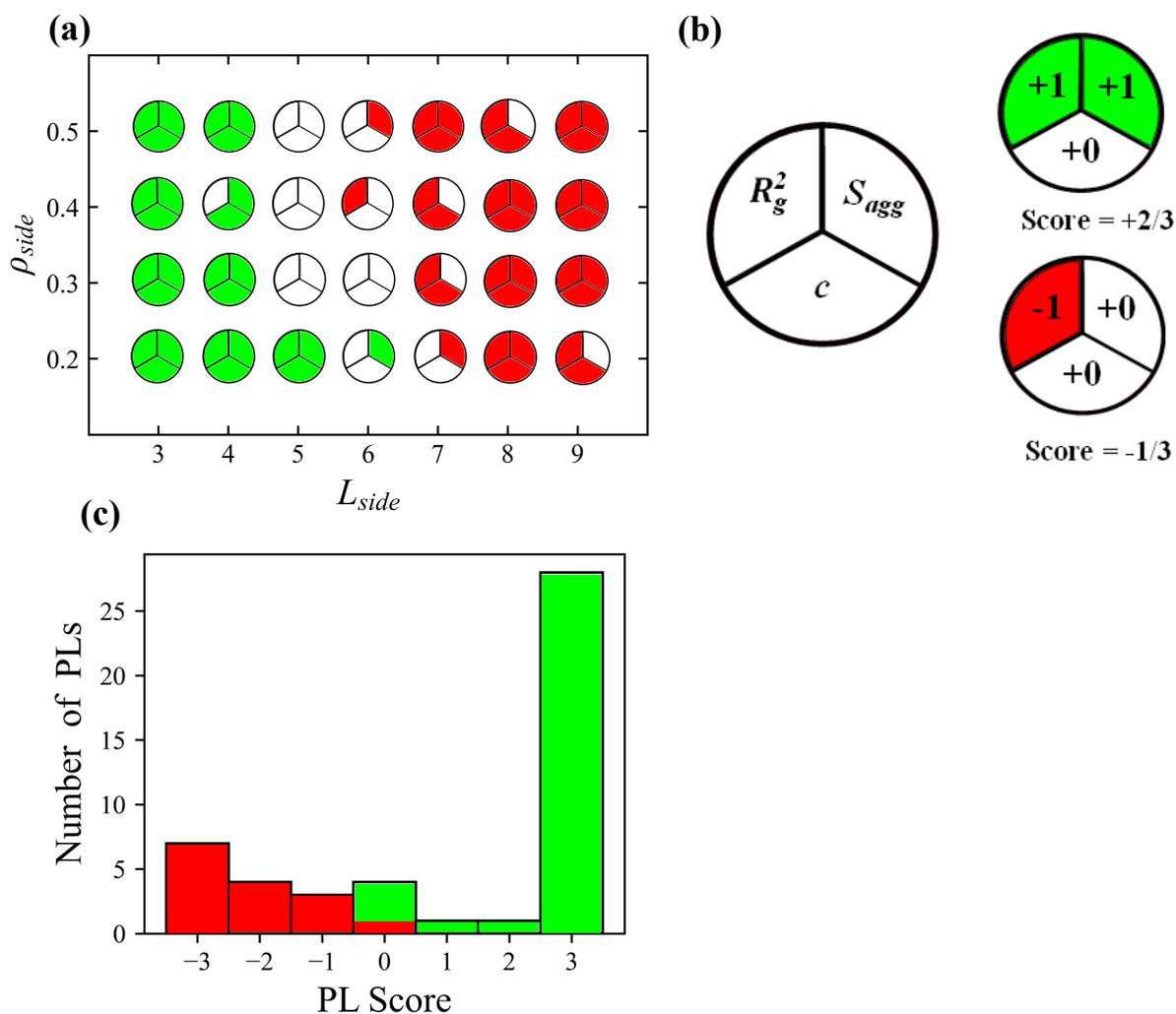

**Figure 8**: (a) PL side chain grafting density against side chain length for rod-like PL topologies. Each point is split into 3 sections that are colored according to whether a PL descriptor successfully classifies a PL as miscible (green) or immiscible (red). Sections that are left blank signify PLs for which the corresponding descriptor cannot accurately determine their behaviour. Each PL is then assigned a corresponding score, displayed in (b), which describes how accurately their behaviour can be predicted by the descriptors. For example, a PL whose behaviour cannot be predicted by any descriptor is awarded a score of +0/3. (c) Histogram displaying the number of PL topologies that received each score.

Both the PL acylindricity, $c$, and its square radius of gyration, $R_g^2$, have a success rate of correctly identifying PLs as miscible or immiscible of 40/48. These two geometrical descriptors, which are computed with the fastest simulations in this work, are not predictors for a mutual set of PL topologies (see 10B-4S-0.4-r and 10B-6S-0.4-r in Table 1), but used in conjunction they are able to exclude 41/48 of the PLs from further simulations to determine PL miscibility. Through the more computationally expensive descriptor, $S_{agg}$, a further 3 PL topologies can be excluded; reducing the test space by a total of 44/48 PLs. However, as it was the case with the geometric descriptors, the aggregation entropy does not exclude a mutual set of PLs with the previous screening steps. This implies that our set of descriptors form a composite picture of the behaviour of the system. Despite this, using the aggregation entropy analysis only on PL topologies which are not filtered by the other descriptors can still be used as an effective layer of the procedure, and indeed the aggregation entropy is the second node split in the decision tree of Figure 6. Using this model to select the miscibility thresholds, i.e. instead of evaluating each descriptor individually, allows us to create a more precise screening procedure based on the descriptors which provide the best performance measures. This decision tree, when built on an expanded dataset, can be used when screening new PLs. If the miscibility cannot conclusively be predicted, full PL/PI simulations or experiments need to be performed. Finally, we note that all descriptors report more accurate results for flexible PLs when compared to rod-like PLs and, among rod-like molecules, those of the lowest and highest molecular weights (see Figure 8(a)) are classified most accurately. This provides useful insight into the PL features which are most likely to be linked to successful miscibility screening.

## Summary and Conclusions

In summary, in this work we have showed that it is possible to use relatively computationally inexpensive molecular simulations to assess the miscibility behaviour of plasticizers dissolved into a polymer matrix by identifying suitable molecular descriptors. The descriptors are decided using a dataset of 48 plasticizer (PL) topologies varying the values of PL side chain length ($L_{side}$), side chain density ($\rho_{side}$) and flexibility and are then used to classify the molecules as miscible or immiscible. They are comprised of both geometrical features of the PL molecules, along with a descriptor of PL configurational entropy which, although an incomplete description of entropy, scales with miscibility and is therefore sufficient for the scope of this work.

Despite the limited size of the dataset, we proved that a supervised learning method such as the decision tree can be used to identify the thresholds for the classification analysis which can therefore be completely automated once a much larger dataset is available. This circumvents the need to assess the performance of the descriptors individually by manually observing the data and can provide more precise thresholds. It also performs feature selection which ensures we collect data with the minimum dimensionality required to accurately classify PL miscibility. The PL topologies display complex miscibility behaviour with some falling in a meta-stable region between mixed and demixed phases. We find that the PL flexibility has a large impact on its miscibility but we demonstrate that PL geometry and size have also an impact. From these results, we can deduce a set of PL design rules. Firstly, the most significant PL feature is the chain flexibility, followed by their topological characteristics. Of these, the first is $\rho_{side}$ which hampers PL miscibility as it increases. The second is $L_{side}$ which has a similar importance. We additionally find that there is a cooperative behaviour between these two properties such that a high value of $\rho_{side}$ may not necessarily produce an immiscible PL/PI system, provided that $L_{side}$ is small and vice versa. This behaviour is limited to the case where the side chain and backbone flexibility are the same, as we have not considered the case in which these factors vary. Finally, we note from preliminary investigations that the screening procedure works also when the backbone length is modified (i.e. shortened or lengthened). We envision that upon the implementation of a high thoughput procedure with a much larger PL set, we will be able to use a decision tree model to construct even more accurate thresholds based on extended data.

Using this procedure, we have reduced the test space by up to 44/48, which greatly reduces the computational cost when testing a large number of PLs. Thus, the workflow can be used to pre-determine PL features which produce target effects on a polymer matrix, and quickly explore PL features which have little prior research in the literature or to reduce the experimental and environmental costs associated to the current way such molecules are chosen. While the current descriptors have an excellent predictivity for miscibility driven by topological effects, if more chemical specific features of the PL molecules affect the solubility, as for example hydrogen bonds, then other descriptors might be needed to be identified. One might envisage a multiscale workflow where an initial screening is performed based only on the topological characteristics of the plasticizers and a further screening, for example based on the enthalpy of mixing, would be carried out on a subset of the already screened samples. Thus, the approach followed here to identify descriptors can be used to develop efficient screening procedures which can be applied to the results of high throughput simulations for other polymeric systems.

**Supporting Information.** Details on the PI/PL systems simulated, PI/vacuum descriptor tests, PL cluster equilibration and aggregation entropy equilibration can be found in the Supporting Information.

**Acknowledgments.** Computational facilities for this work were provided by the Computational Shared Facility (CSF) of the University of Manchester. LS acknowledges financial support from EPSRC CDT

Graphene NOWNANO (Grant No. EP/L01548X/1) and Continental Tires. AT gratefully acknowledges the support from the European Research Council (Grant No. 101020369).

# A Framework for a High Throughput Screening Method to Assess Polymer/Plasticizer Miscibility

*Lois Smith[a], H. Ali Karimi-Varzaneh[b], Sebastian Finger[b], G. Giunta,[a,c] A. Troisi[d], Paola Carbone[*a]*

[a] *Department of Chemical Engineering, School of Engineering, The University of Manchester, Oxford Road, M13 9PL, Manchester, United Kingdom*

[b] *Continental Reifen Deutschland GmbH, Jädekamp 30, D-30419 Hanover, Germany*

[c] *BASF, Carl-Bosch-Strasse 38, 67056, Ludwigshafen, Germany*

[d] *Department of Chemistry, Department of Chemistry, Crown Street, L69 7ZD, Liverpool, United Kingdom*

[*] *paola.carbone@manchester.ac.uk*

*List of Systems Simulated and Miscibility Parameters*

**Table S1** List of all PL topologies simulated in this work, number of PL molecules in a system with 72 PI chains of molecular weight 300, the simulation box size after NPT equilibration and corresponding miscibility parameter, ζ.

| Topology Number | Code | PL Molecule Number | Box Size (nm) | Miscibility Parameter, ζ |
|---|---|---|---|---|
| 1 | 10B-3S-0.2-r | 68 | 8.654 × 8.654 × 25.229 | 1.27 ± 0.004 |
| 2 | 10B-3S-0.3-r | 57 | 8.656 × 8.656 × 25.236 | 1.32 ± 0.003 |
| 3 | 10B-3S-0.4-r | 49 | 8.656 × 8.656 × 25.235 | 1.46 ± 0.01 |
| 4 | 10B-3S-0.5-r | 43 | 8.653 × 8.653 × 25.227 | 1.40 ± 0.005 |
| 5 | 10B-3S-0.2-f | 68 | 8.654 × 8.654 × 25.229 | 1.25 ± 0.003 |
| 6 | 10B-3S-0.3-f | 57 | 8.656 × 8.656 × 25.240 | 1.32 ± 0.005 |
| 7 | 10B-3S-0.4-f | 49 | 8.656 × 8.656 × 25.235 | 1.33 ± 0.004 |
| 8 | 10B-3S-0.5-f | 43 | 8.653 × 8.653 × 25.227 | 1.39 ± 0.005 |
| 9 | 10B-4S-0.2-r | 60 | 8.655 × 8.655 × 25.233 | 1.28 ± 0.003 |
| 10 | 10B-4S-0.3-r | 49 | 8.655 × 8.655 × 25.233 | 1.43 ± 0.01 |
| 11 | 10B-4S-0.4-r | 42 | 8.657 × 8.657 × 25.240 | 1.47 ± 0.01 |
| 12 | 10B-4S-0.5-r | 36 | 8.654 × 8.654 × 25.231 | 1.75 ± 0.01 |
| 13 | 10B-4S-0.2-f | 60 | 8.655 × 8.655 × 25.233 | 1.28 ± 0.004 |
| 14 | 10B-4S-0.3-f | 49 | 8.655 × 8.655 × 25.233 | 1.34 ± 0.004 |
| 15 | 10B-4S-0.4-f | 42 | 8.657 × 8.657 × 25.240 | 1.39 ± 0.004 |
| 16 | 10B-4S-0.5-f | 36 | 8.654 × 8.654 × 25.231 | 1.47 ± 0.004 |
| 17 | 10B-5S-0.2-r | 54 | 8.735 × 8.735 × 24.884 | 1.39 ± 0.01 |
| 18 | 10B-5S-0.3-r | 43 | 8.143 × 8.143 × 28.490 | 1.56 ± 0.01 |
| 19 | 10B-5S-0.4-r | 36 | 8.720 × 8.720 × 24.841 | 1.68 ± 0.01 |
| 20 | 10B-5S-0.5-r | 31 | 8.720 × 8.720 × 24.839 | 48.25 ± 0.1 |
| 21 | 10B-5S-0.2-f | 54 | 8.735 × 8.735 × 24.884 | 1.32 ± 0.005 |
| 22 | 10B-5S-0.3-f | 43 | 8.143 × 8.143 × 28.490 | 1.39 ± 0.01 |
| 23 | 10B-5S-0.4-f | 36 | 8.720 × 8.720 × 24.841 | 1.50 ± 0.01 |
| 24 | 10B-5S-0.5-f | 31 | 8.720 × 8.720 × 24.839 | 1.62 ± 0.01 |
| 25 | 10B-6S-0.2-r | 49 | 8.724 × 8.724 × 24.851 | 1.57 ± 0.01 |
| 26 | 10B-6S-0.3-r | 39 | 8.653 × 8.653 × 25.228 | 1.75 ± 0.01 |
| 27 | 10B-6S-0.4-r | 32 | 8.659 × 8.659 × 25.244 | 39.44 ± 0.21 |
| 28 | 10B-6S-0.5-r | 27 | 8.652 × 8.652 × 25.226 | 69.87 ± 0.20 |
| 29 | 10B-6S-0.2-f | 49 | 8.724 × 8.724 × 24.851 | 1.28 ± 0.005 |
| 30 | 10B-6S-0.3-f | 39 | 8.653 × 8.653 × 25.228 | 1.43 ± 0.01 |
| 31 | 10B-6S-0.4-f | 32 | 8.659 × 8.659 × 25.244 | 1.54 ± 0.01 |
| 32 | 10B-6S-0.5-f | 27 | 8.652 × 8.652 × 25.226 | 1.72 ± 0.01 |
| 33 | 10B-7S-0.2-r | 45 | 8.654 × 8.654 × 25.230 | 19.43 ± 0.04 |
| 34 | 10B-7S-0.3-r | 35 | 8.653 × 8.653 × 25.229 | 25.23 ± 0.1 |
| 35 | 10B-7S-0.4-r | 28 | 8.720 × 8.720 × 24.839 | 16.60 ± 0.02 |
| 36 | 10B-7S-0.5-r | 24 | 8.654 × 8.654 × 25.229 | 17.04 ± 0.01 |
| 37 | 10B-7S-0.2-f | 45 | 8.654 × 8.654 × 25.230 | 1.38 ± 0.01 |
| 38 | 10B-7S-0.3-f | 35 | 8.654 × 8.654 × 25.230 | 1.51 ± 0.01 |
| 39 | 10B-7S-0.4-f | 28 | 8.720 × 8.720 × 24.839 | 1.64 ± 0.01 |
| 40 | 10B-7S-0.5-f | 24 | 8.654 × 8.654 × 25.229 | 1.68 ± 0.01 |
| 41 | 10B-8S-0.2-r | 42 | 8.144 × 8.144 × 28.493 | 12.80 ± 0.03 |
| 42 | 10B-8S-0.3-r | 32 | 8.145 × 8.145 × 28.495 | 64.83 ± 0.21 |
| 43 | 10B-8S-0.4-r | 26 | 8.145 × 8.145 × 28.494 | 15.66 ± 0.01 |
| 44 | 10B-8S-0.5-r | 22 | 8.143 × 8.143 × 28.490 | 37.16 ± 0.08 |
| 45 | 10B-9S-0.2-r | 39 | 8.145 × 8.145 × 28.495 | 11.86 ± 0.02 |
| 46 | 10B-9S-0.3-r | 29 | 8.143 × 8.143 × 28.490 | 79.41 ± 0.31 |
| 47 | 10B-9S-0.4-r | 23 | 8.140 × 8.140 × 28.477 | 31.78 ± 0.11 |
| 48 | 10B-9S-0.5-r | 20 | 8.146 × 8.146 × 28.499 | 10.77 ± 0.02 |

*Vacuum Simulation Validation*

In order to confirm the validity of calculating PL molecule square radius of gyration, $R_g^2$, acylindricity, $c$, and configurational entropy, $S_{config}$, with a simulation of a single PL in vacuum, we compared these analyses to those done from a system of a PL immersed in a PI melt. To do so, we chose a small subset of 8 PLs, composed of both flexible and rod-like molecules. Fig.

S3 and S4 display the results. All results are extracted over approximately 3 μs and the PL/PI simulations undergo an initial NPT equilibration, with the simulation parameters disclosed in the main text, to achieve an appropriate density. To ensure the PI melt is then properly equilibrated, a further 800 ns NVT simulation is performed which is sufficient time for the PI mean square internal distance (MSID) to plateau.

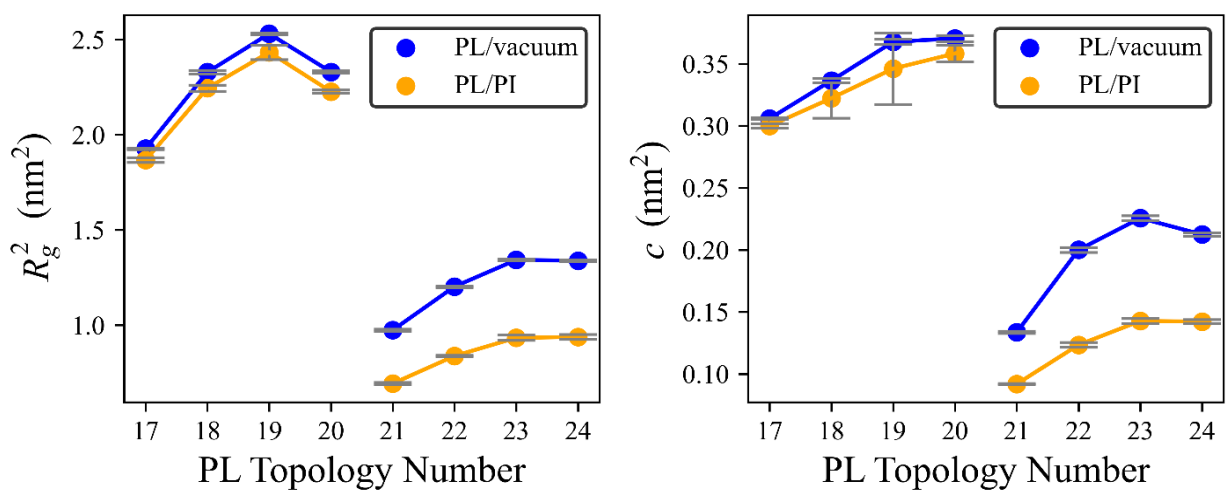

**Figure S1**: $R_g^2$ (left) and $c$ (right) for 8 PL topologies as labelled. Values are taken from (blue) a PL molecule in an empty box and (orange) a PL molecule in an equilibrated PI melt. All results are extracted over a simulation of length 2 μs. The error bars are obtained with the standard error of block averages.

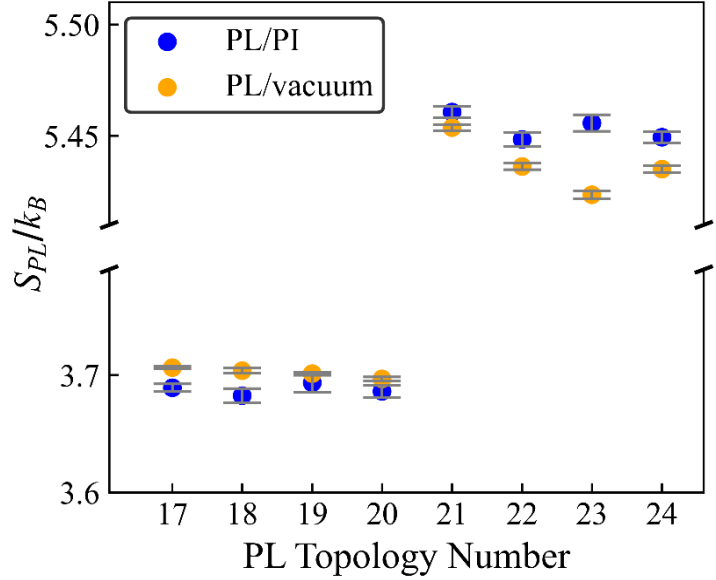

**Figure S2**: Configurational entropy, $S_{PL}$, for 8 PL topologies as labelled. Values are taken from (blue) a PL molecule in an empty box and (orange) a PL molecule in an equilibrated PI melt. All results are extracted over a simulation of length 2 μs. The error bars are obtained with the standard error of the data.

Figure S1 displays a comparison between (a) $R_g^2$ and (b) $c$ in vacuum and in a PI melt. In both cases, a more significant difference is seen between the systems with flexible PLs (topologies 21, 22, 23 and 24), however the trend is similar. This is likely due to a swelling of the flexible PL molecules in vacuum, which is observed to a lesser extent in the rod-like molecules due to the rigid nature of their bond angles. Figure S2 displays the results for the PL configurational entropy, $S_{PL}$, for which we observe a slight but not significant, difference in results between the PI and vacuum systems.

The results for this PL subsets imply it is valid to develop the simulation procedure making use of simulating PL molecules in vacuum. This enables us to significantly reduce computational cost.

*PL Miscibility*

The following section displays example snapshots of miscible and immiscible PL topologies in systems of PL molecules immersed in a PI melt using the bead and spring model described in the main text. The concentration of PL molecules is fixed at approximately 5 *phr*. The PI melt has been removed from the visualisation. Figure S4(a) displays a system where $\zeta \approx 1.5$, which is typical of a system that displays miscible PL behaviour. In contrast, S4(b) shows a system with partial PL clusters formed and $\zeta \approx 2.7$, which we conservatively choose as a cut-off value to distinguish between miscible and immiscible PLs. The third image, S4(c), shows a system which has formed large clusters and, as such, has a significantly larger $\zeta$ value, $\zeta \approx 69.70$. Figure S5 shows an example of miscible (red) and immiscible (blue) PL behaviour in terms of the miscibility parameter, $\zeta$. Both systems begin with PL molecules evenly dispersed throughout the simulation box. PL 10B-3S-0.2-r (miscible) remains fluctuating around a value of $\zeta = 1.7$ across the trajectory, whereas PL 10B-7S-0.4-r (immiscible) shows a rising value of $\zeta$ over approximately 5 μs which corresponds to an increase in PL clustering.

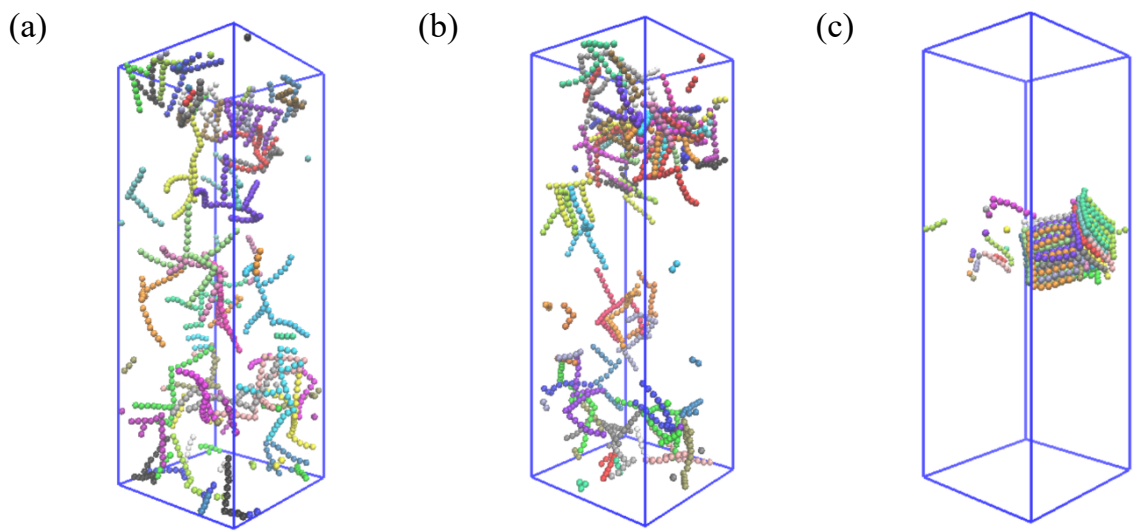

**Figure S4**: Snapshots of PL molecules in the PL/PI (PI not pictured) systems. The PLs depicted are **(a)** 10B-7S-0.3-r, **(b)** 10B-7S-0.4-r and **(c)** 10B-6S-0.5-r. The approximate $\zeta$ values are **(a)** 1.5, **(b)** 2.7 and **(c)** 69.70. Each PL molecule is independently coloured. From this, an increase in PL agglomeration can be observed from **(a)-(c)**.

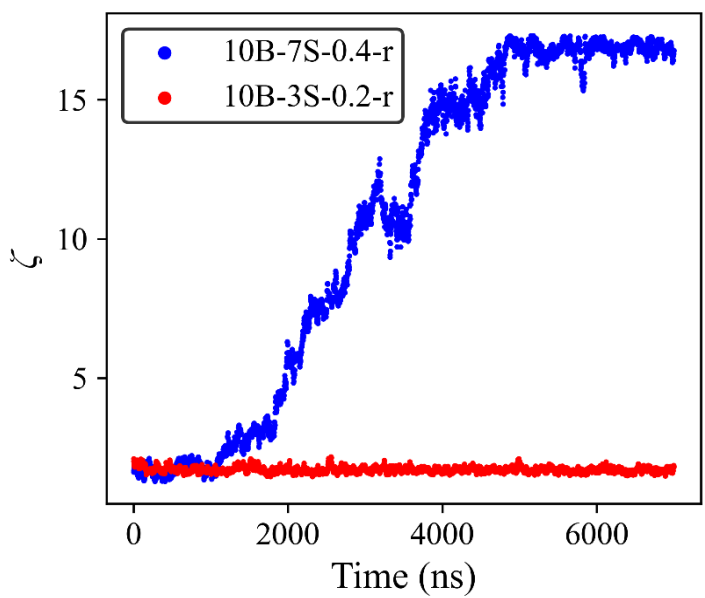

**Figure S5**: Miscibility factor, $\zeta$, for PL 10B-3S-0.2-f (red) and 10B-7S-0.4-r (blue) against simulation time.

*PL/PL System Equilibration*

The equilibration of the PL/PL simulations which are used to calculate the PL configurational entropy of PLs in infinite clusters, $S_{PL/PL}$, is identified by calculating the probability density associated with Equation (11) with the time averaged PL end-to-end distance distribution over 600 ns blocks. We also average over each PL in the system for good statistics. We find this is sufficient time for $S_{PL/PL}$ to reach a plateau, which takes approximately 8 μs for the systems with rod-like PLs, as displayed in Figure S6. The other, flexible PL, systems have a significantly shorter equilibration period due to the less rigid nature of the constraints on their bond angles.

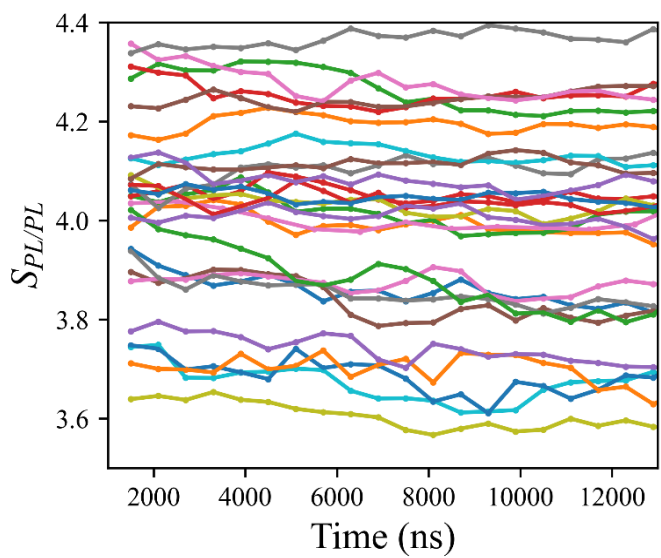

**Figure S6**: Plot showing trends in the equilibration periods of PL/PL simulations with rod-like PL topologies. Each colour represents a different PL topology and results for $S_{PL/PL}$ are extracted after an 8 μs time period.

*Further Flexible PL Simulations*

We performed simulations to calculate each descriptor for a further 3 flexible PLs; in order to verify that our procedure is unaffected by the placement of side chains on the first or last beads of the PL 'backbone', such that the molecules retain a more 'brush-like' shape comparable with their rod-like counterparts. The PLs simulated are depicted in Figure S7 and corresponding results for the $R_g^2$, $c$ and $S_{agg}$ descriptors are displayed in Figure S8.

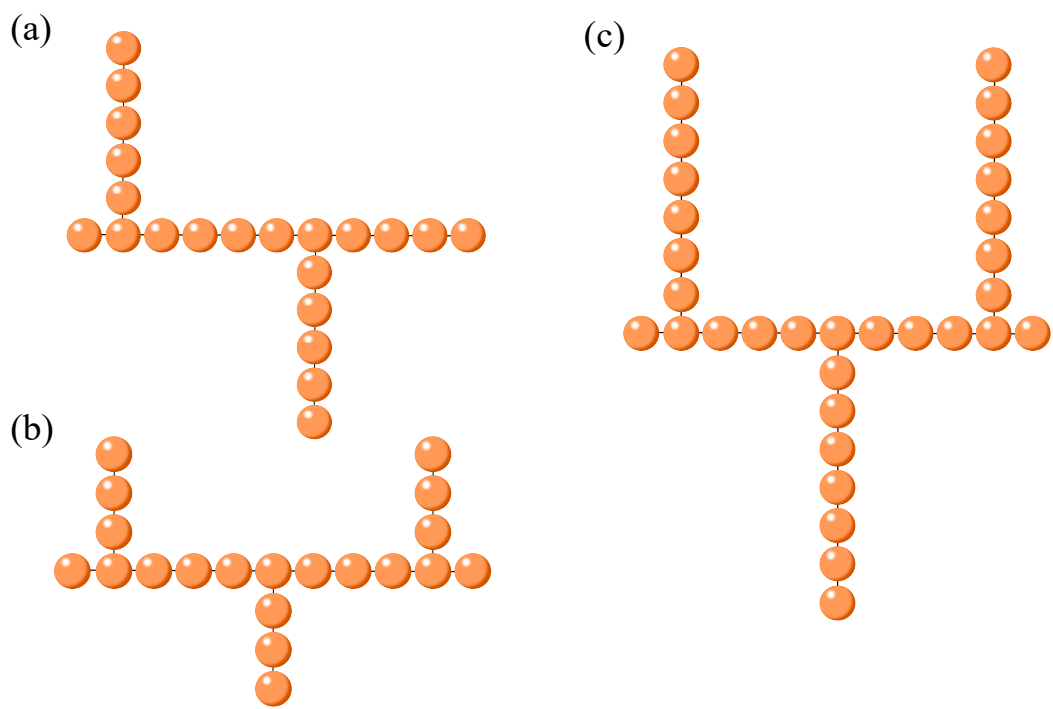

**Figure S7**: Schematic representations of the 3 additional flexible PLs simulated. Each backbone contains 11 beads and side chain lengths and frequencies were chosen to approximately reproduce the 'brush-like' shape present in the rod-like PLs displayed in **Figure 1(a)**, **(c)** and **(f)** of the main text.

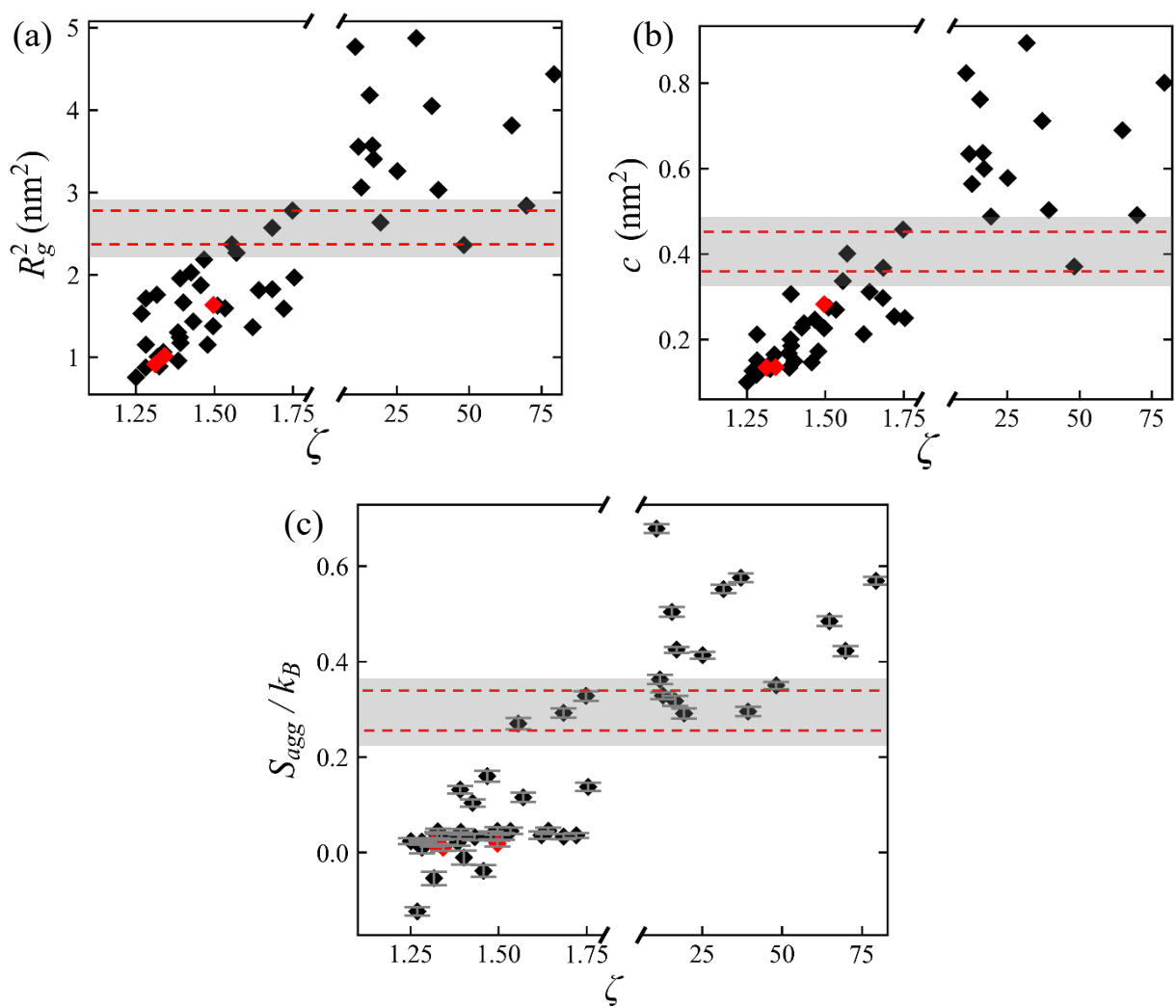

**Figure S8**: Plots of **(a)** $R_g^2$, **(b)** $c$ and **(c)** $S_{agg}$ against miscibility parameter, $\zeta$, for the 48 PLs in this work, along with those displayed in Figure S8 (red).

The results are in line with our previous findings that flexible PLs are miscible within the PI matrix and their descriptor values correlate with $\zeta$ accordingly, regardless of the placement of their side chains.